\documentclass[12pt]{iopart}
\usepackage[
colorlinks=true,
frenchlinks=false,
linkcolor=blue,
anchorcolor=blue,
citecolor=blue,
filecolor=blue,
urlcolor=blue,
bookmarks=true,
bookmarksopen=true,
bookmarksnumbered=true,
bookmarksopenlevel=1,
plainpages=false,
pdfpagelabels=true,
breaklinks
]{hyperref}
\usepackage{cite}
\usepackage{graphicx}
\usepackage{dcolumn}
\usepackage{bm}
\usepackage{gensymb} 
\usepackage{lineno}
\expandafter\let\csname equation*\endcsname=\relax
\expandafter\let\csname endequation*\endcsname=\relax
\usepackage{amsmath}
\usepackage{color}
\usepackage[utf8]{inputenc}
\usepackage[T1]{fontenc}
\usepackage{mathptmx}
\usepackage{array, multirow, bigdelim, makecell, booktabs} 

\def\SymbReg{\textsuperscript{\textregistered}}
\begin{document}
\title[Magnetic field robust high quality factor NbTiN superconducting microwave resonators]{Magnetic field robust high quality factor NbTiN superconducting microwave resonators}
\author{M.~M\"uller$^1$ $^2$, T.~Luschmann$^1$ $^2$ $^3$, A.~Faltermeier$^1$ $^2$, S.~Weichselbaumer$^1$ $^2$ $^3$, L.~Koch$^1$ $^2$ $^3$, G.B.P.~Huber$^1$ $^2$, H.W.~Schumacher$^4$, N.~Ubbelohde$^4$, D.~Reifert$^4$, T.~Scheller$^4$, F.~Deppe$^1$ $^2$ $^3$, A.~Marx$^1$, S.~Filipp$^1$ $^2$ $^3$, M.~Althammer$^1$ $^2$, R.~Gross$^1$ $^2$ $^3$ and H.~Huebl$^1$ $^2$ $^3$ }
\address{$^1$ Walther-Mei{\ss}ner-Institut, Bayerische Akademie der Wissenschaften, 85748 Garching, Germany}
\address{$^2$ Physik-Department, Technische Universit\"{a}t M\"{u}nchen, 85748 Garching, Germany}

\address{$^3$ Munich Center for Quantum Science and Technology (MCQST), 80799 Munich, Germany}
\address{$^4$ Physikalisch-Technische Bundesanstalt, 38112 Braunschweig, Germany}

\eads{\mailto{manuel.mueller@wmi.badw.de}}
\date{\today}
\begin{abstract}
\noindent We systematically study the performance of compact lumped element planar microwave $\mathrm{Nb_{70}Ti_{30}N}$ (NbTiN) resonators operating at 5 GHz in external in-plane magnetic fields up to 440 mT, a broad temperature regime from 2.2 K up to 13 K, as well as mK temperatures. For comparison, the resonators have been fabricated on thermally oxidized and pristine, (001) oriented silicon substrates. When operating the resonators in the multi-photon regime at $T=2.2$ K, we find internal quality factors $Q_{\mathrm{int}}\simeq$ $2\cdot10^5$ for NbTiN resonators grown on pristine Si substrates, while resonators grown on thermally oxidized substrates show a reduced value of $Q_{\mathrm{int}}\simeq$ $1\cdot10^4$, providing evidence for additional loss channels for the latter substrate. In addition, we investigate 
the $Q$-factors of the resonators on pristine Si substrates at millikelvin temperatures to asses their applicability for quantum applications. We find $Q_{\mathrm{int}}\simeq$ $2\cdot10^5$ in the single photon regime and $Q_{\mathrm{int}}\simeq$ $5\cdot10^5$ in the high power regime at $T=7$ mK.
\end{abstract}
\noindent{\it Superconducting planar microwave resonators, niobium titanium nitride thin films, electron spin resonance, ferromagnetic resonance, microwave resonators for circuit QED, performance of microwave resonators at millikelvin temperatures, dc-sputter deposition of niobium titanium nitride }
\maketitle
\section{Introduction}
High quality superconducting microwave resonators are central building blocks for today's quantum science and technology. This particularly includes quantum information processing with superconducting circuit elements \cite{Blais2021, AndreasWallraff2004, Schoelkopf2008, Wallraff2004, Koch2007} and the realization of photon detectors \cite{Korneeva2018, Dorenbos2011, Kaniber2016, Reithmaier2015, Reithmaier2014, Reithmaier2013}. Moreover, they are considered an important enabling technology for a variety of quantum devices such as quantum limited microwave amplifiers \cite{Bergeal2010,Castellanos-Beltran2007,Tholen2007}. They also play a key role in hybrid quantum systems like nano-electromechanical systems \cite{Aspelmeyer2014}, cavity magnonic systems \cite{Li2019a, Tabuchi2014, Tabuchi2015, Huebl2013,Hou2019} and hybrids based on the spin-photon interaction \cite{Schuster2010,Kubo2011,Majer2007, Weichselbaumer2020, Zollitsch2015, Bushev2011, Probst2014}. A particular advantage of microwave resonators based on superconductors is the fact that resonators with high and ultra-high quality factors $Q$ can be realized even for planar designs, e.g. in the form of lumped-element or coplanar waveguide resonators\cite{Zollitsch2015,Narkowicz2005,Wallace1991}. While aluminum and niobium represent the well-established materials in superconducting quantum technology, superconductors with higher critical temperatures $T_{\mathrm{c}}$ exist and are discussed in this context. Niobium titanium nitride (NbTiN) is one of the prime candidates, as it is a hard type II superconductor with a transition temperature of up to $T_\mathrm{c}=17$ K \cite{DiLeo1990, Horn1968, Systems2004} and a large Ginzburg-Landau parameter $\kappa_{\mathrm{GL}} > 50$ ($\kappa_{\mathrm{GL}}=\lambda_{\mathrm{L}}/\xi_{\mathrm{GL}}$, where $\lambda_{\mathrm{L}}$ is the London penetration depth and $\xi_{\mathrm{GL}}$ is the superconducting Ginzburg-Landau coherence length. For NbTiN, we assume $\lambda_{\mathrm{L}}^{\mathrm{NbTiN}}=200$ nm \cite{Yu2005} and $\xi_{\mathrm{GL}}^{\mathrm{NbTiN}}=3.8$ nm \cite{Yu2002}). The higher $T_{\mathrm{c}}$ typically translates into a larger superconducting gap and the large $\kappa_{\mathrm{GL}}$ into a high upper critical field. Altogether, this offers an enhanced robustness against magnetic fields and better performance at 'elevated' temperatures due to the correspondingly lower density of thermally induced quasiparticles. These properties are of interest for pushing hybrid quantum systems to elevated temperatures and are discussed in the context of high and ultra-high frequency superconducting qubits. Here, we investigate the performance of thin-film $\mathrm{Nb_{70}Ti_{30}N}$ (NbTiN) lumped element microwave resonators from a materials perspective. Specifically, we focus on the preparation of the NbTiN thin-films and compare the achieved quality factors for various substrate configurations.
\section{Experimental details}
\begin{figure}[htpb]
	\centering
	\includegraphics[scale=1.0]{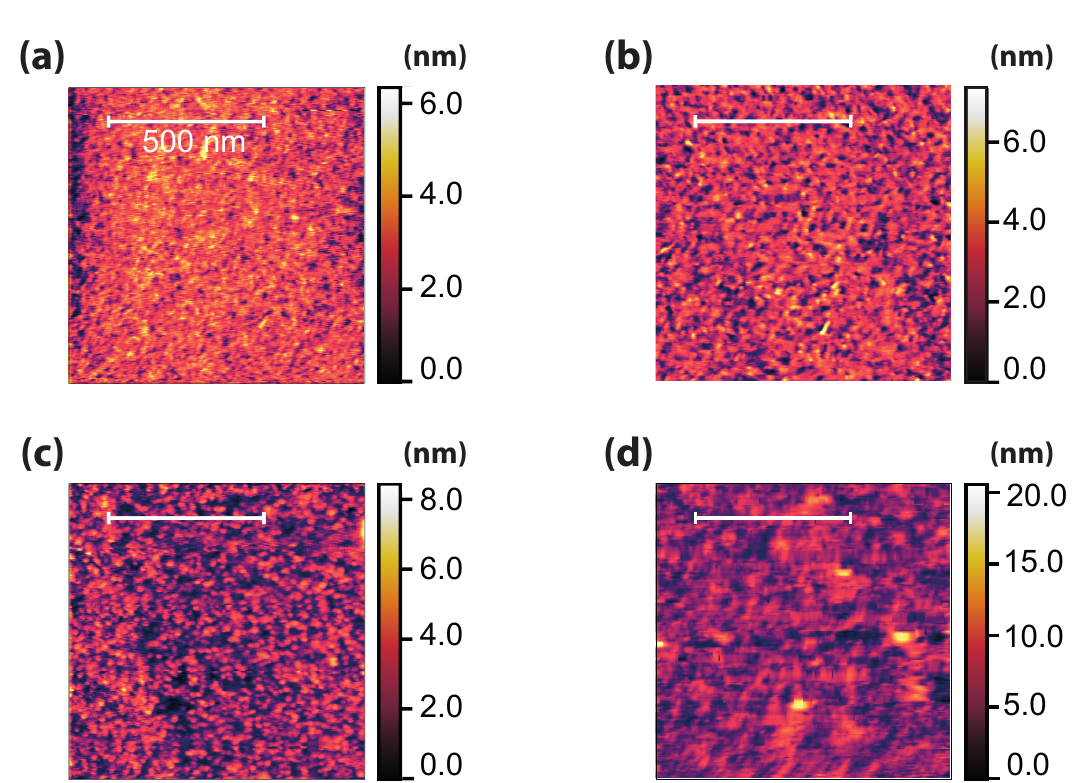}
	\caption{AFM surface scans of a $1 \mathrm{\mu m^2}$-area for the NbTiN film deposited on (a) a $\mathrm{SiO_2}$-substrate (b) Si-substrate, (c) a $\mathrm{SiO_2}$-substrate from the PTB, (d) a HF-dipped Si-substrate. The extracted RMS surface roughnesses are (a) 0.6 nm, (b) 0.7 nm, (c) 0.8 nm and (d) 1.5 nm, respectively.}
	\label{Fig: series1}
\end{figure}

\begin{figure}[htpb]
	\centering
	\includegraphics[scale=1.0]{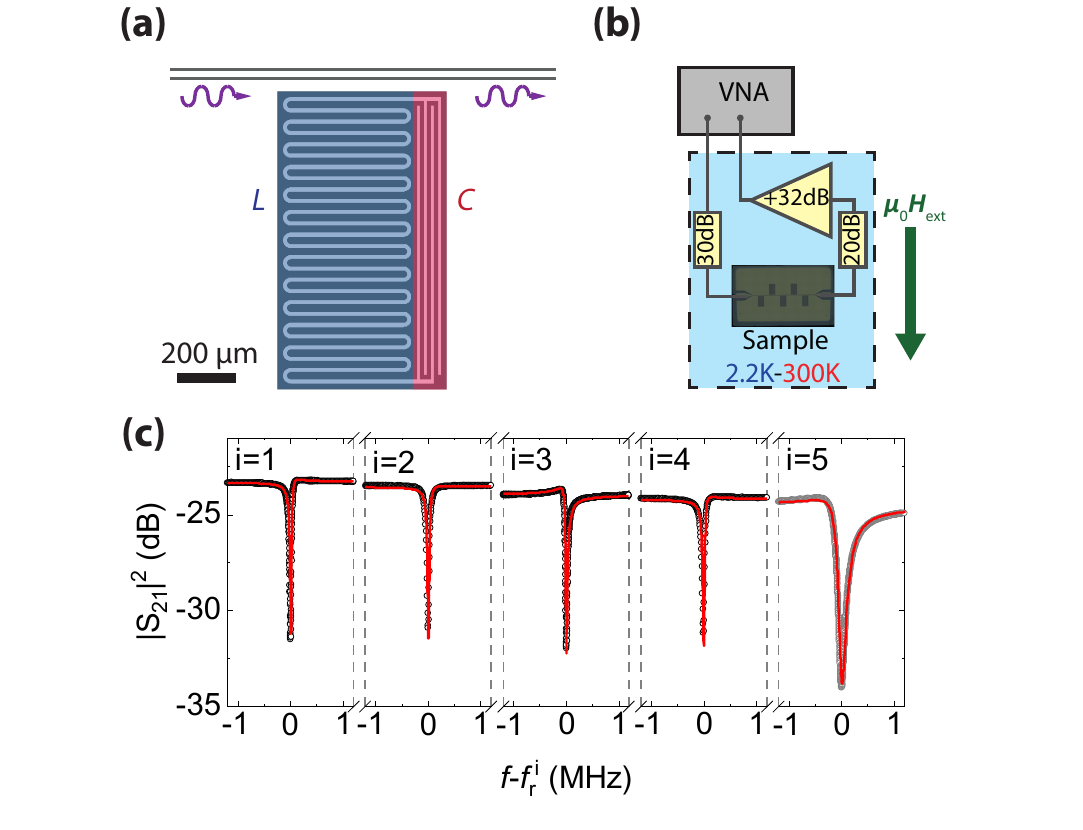}
	\caption{(a) Schematic illustration of the resonator design studied in this work. The capacitively shunted meander-type resonator (CR) layout consists of an interdigitated finger capacitor (red) shunting a meandering inductor (blue). (b) Schematic of the microwave circuit of the measurement setup. The green arrow indicates the direction of the applied in-plane magnetic field $\mu_0H_{\mathrm{ext}}$. (c) Exemplary microwave transmission spectrum $|S_{21}(f)|^2$ of a device fabricated from NbTiN grown on Si, measured at $T=2.2$ K and at a microwave power corresponding to  $\simeq10^7$ photons on average in the resonator. One can identify five resonance features, which can be attributed to the five resonators (R1-R5) patterned into the chip. Fits to the data according to Eq.\,(1) are shown as red lines. Each of the panels has a frequency width of 2 MHz. The data points of resonator R5 are colored gray as they are omitted from the statistical analysis. The fitting parameters for each of the resonators are summarized in Tab.\,\ref{Tab: 1}.
	}
	\label{Fig: series2}
\end{figure}

The superconducting resonators are patterned into NbTiN thin films grown by sputter deposition on both pristine (001) oriented high-resistivity Si substrates ($\rho>10^4 \;\Omega$cm) and Si substrates covered by thermal oxide layers of 1 $\mu$m thickness ($\rho>4\cdot10^3 \;\Omega$cm). We further compare the characteristic properties of the resonators fabricated on these substrates with those achieved for resonators with identical geometry fabricated from NbTiN thin films deposited on a thermally oxidized Si substrate at the \textit{Physikalisch-Technischen Bundesanstalt} (PTB). In addition to the low temperature limit, which is of importance for applications in superconducting quantum technology, we investigate the performance of our resonators at temperatures up to $T=13$ K and at applied magnetic fields up to $\mu_0H_{\mathrm{ext}}=440$ mT. The latter parameter regime is particularly relevant for applications such as electron spin resonance (ESR) or ferromagnetic resonance (FMR).\\
At the Walther-Mei{\ss}ner-Institute (WMI), the $\mathrm{Nb_{70}Ti_{30}N}$ thin films with a layer thickness of $d=150$ nm were grown on both thermally oxidized and bare Si (100) substrates using reactive dc magnetron sputtering in a mixed Ar/$\mathrm{N_2}$-atmosphere with a gas flow ratio of 36.2/3.8, a deposition temperature of $T_{\mathrm{depo}}=500\,^\circ$C, a pressure of $p_{\mathrm{depo}}=5\mathrm{\;\mu}$bar, and a deposition power of $P_{\mathrm{s}}=95$ W using a 4" $\mathrm{Nb_{70}Ti_{30}}$ target. These sputtering parameters are the result of a growth optimization series to maximize the superconducting transition temperature of NbTiN reaching a maximum of $T_{\mathrm{c}}=16.3$ K, determined by dc resistance measurements.\\
At the PTB, the NbTiN was grown with a layer thickness of 150 nm on a 300 nm thick thermally oxidized Si substrate with a resistivity of 1-10$\;\Omega$cm at RT at an Ar/$\mathrm{N_2}$ gas flow ratio of 20/2.6 without any substrate heating and thermalization, a pressure of $p_{\mathrm{depo}}=5\mathrm{\;\mu}$bar and at a deposition power of $P_{\mathrm{s}}=320$ W using a 6" $\mathrm{Nb_{70}Ti_{30}}$ target. \\
Atomic force microscopy (AFM) surface roughness scans shown in Fig.\,\ref{Fig: series1} indicate low root mean squared (RMS) surface roughness values of less than 1 nm for the NbTiN films grown on (a) a $\mathrm{SiO_2}$ substrate ($\mathrm{RMS}=0.6$ nm) at the WMI, (b) a Si substrate ($\mathrm{RMS}=0.7$ nm) grown at the WMI and (c) the NbTiN thin films prepared at the PTB on a $\mathrm{SiO_2}$ substrate ($\mathrm{RMS}=0.8$ nm). Only the NbTiN film grown on a Si substrate at the WMI where the native oxide was stripped of prior to the deposition using a HF dip (d) exhibits a higher roughness ($\mathrm{RMS}=1.5$ nm) and larger visible grains. We attribute this to the highly textured growth of NbTiN induced by the removal of the native amorphous oxide layer. We note that only this NbTiN thin film exhibited crystalline reflexes in x-ray diffraction scans which can be attributed to a highly textured film with the (002) plane of the cubic phase oriented parallel to the substrate surface \cite{Ge2019,Zhang2015,Myoren2001}. A study of the crystalline reflexes of this sample has been added to the supplemental material (SM) section I \cite{Supplements}.
Generally, a low surface roughness is desirable to reduce losses in the planar microwave resonators at the metal-air (MA) interface \cite{Megrant2012}.\\ 
The NbTiN films are patterned into planar lumped element resonators using electron beam lithography and reactive ion etching (RIE) in a mixed $\mathrm{SF_6}$/Ar-atmosphere. The investigated chip layout includes a microwave transmission or feedline, which is coupled in a hanger-type configuration to five lumped element microwave resonators which slightly differ in their capacitance $C$ and hence their resonance frequency $f_\mathrm{r}$. This configuration allows for a multiplexed readout scheme, which we use to asses the resonator to resonator variance with respect to the $Q$-factor [see also Fig.\,\ref{Fig: series2}(a) for the geometry of the resonator layout]. For measurements, the finalized chip is mounted in a gold plated oxygen free high thermal conductivity copper sample box and inserted into a variable temperature helium cryostat.
Fig.\,\ref{Fig: series2}(b) schematically depicts the microwave transmission setup employed for our experiments. We connect the two ends of the central feed line to the two ports of a vector network analyzer (VNA) and record the complex transmission parameter $S_{21}(f)$ close to the resonance frequency $f_r$ for each resonator [cf. Fig.\, \ref{Fig: series2} (c)] at applied microwave powers that corresponds to the high-photon limit of the resonators ($\langle n_{\mathrm{ph}}\rangle\simeq 10^7$). We note that at temperatures in the Kelvin scale, only negligible changes in resonator quality factors with varying $\langle n_{\mathrm{ph}}\rangle$ were observed as the power-dependent two-level systems (TLS) are thermally depolarized at these high $T$.
\begin{table}[htpb]	
	\centering
	\begin{tabular}{|c|c|c|c|c|}
		\hline	
		Resonator	&	$f_{\mathrm{r}}$ (GHz) & $Q$ ($\times 10^3$) & $Q_{\mathrm{int}}$ ($\times 10^3$) & $Q_{\mathrm{ext}}$($\times 10^3$) \\ \hline
		R1	&	4.672 & 84.7$\pm$1.4 & 215.2$\pm$9.0 & 139.7$\pm$0.1 \\ \hline
		R2	&	4.806 &61.0$\pm$0.1 & 152.2$\pm$5.3 &101.9$\pm$0.1 \\ \hline
		R3	&	4.924 &84.3$\pm$0.6 & 207.8$\pm$3.8 &141.9$\pm$0.1 \\ \hline
		R4	&	5.026 &69.4$\pm$2.3 & 161.7$\pm$12.0 &121.5$\pm$0.1 \\ \hline
		R$5^*$	&	5.133 &17.9$\pm$0.1 & 50.6$\pm$0.1 & 27.9$\pm$0.1 \\ \hline
	\end{tabular}
	\caption{Extracted fitting parameters for resonance frequency $f_\mathrm{r}$, quality factor $Q$, internal quality factor $Q_{\mathrm{int}}$ and external quality factor $Q_{\mathrm{ext}}$ for the on chip resonators. Measurements are performed at $T=2.2$ K and at a microwave power corresponding to $\simeq10^7$ photons on average in the resonator. Resonator R5 is marked with a $^*$, as it has been omitted from the statistical analysis due to its exceptionally bad performance.}
	\label{Tab: 1}
\end{table}
To guarantee the thermalization of the inner conductor of the microwave line and to reduce the impact of room temperature thermal noise, the input signal is attenuated by 30 dB close to the SMA connector of the sample box for the experiments in the helium cryostat. For the experiments performed in the dilution refrigerator we place the attenuators at the various temperature stages to achieve a thermal microwave photon population of $n_{\mathrm{ph}}\ll 1$ ( see SM, Fig.\,S3 \cite{Supplements}). At the output connector, we use a 20 dB attenuator for thermalization and noise purposes for the experiments in the Kelvin range as well as a cryogenic HEMT amplifier (gain: +32 dB). By design, each resonator is coupled to the shared feed line with the external loss rate $\kappa_{\mathrm{ext}}$, which is related to the external quality factor $Q_{\mathrm{ext}}=2\pi f_{\mathrm{r}}/2\kappa_{\mathrm{ext}}$. The internal losses are parameterized by the internal loss rate $\kappa_{\mathrm{int}}$ which corresponds to the internal quality factor via $Q_{\mathrm{int}}=2\pi f_{\mathrm{r}}/2\kappa_{\mathrm{int}}$. Using this definition, the half width at half maximum linewidth (HWHM) is given by $\kappa/2\pi=(\kappa_{\mathrm{ext}}+\kappa_{\mathrm{int}})/2\pi$ which relates to the total quality factor via $ Q=2\pi f_{\mathrm{r}}/2\kappa$. The Q-factors can be extracted from the complex transmission data $S_{21}(f)$ using the 'circle fit method' \cite{Probst2015}
\begin{equation}
	S_{21}^{\mathrm{notch}}\left(f\right)=ae^{i\beta}e^{-2\pi i f \tau}\left[1-\frac{\left(Q/|Q_{\mathrm{ext}}|\right)e^{i\phi}}{1+2iQ\left(f/f_{\mathrm{r}}-1\right)}\right].
	\label{Eq: circle-fit}
\end{equation}
The prefactors in front of the square brackets account for the attenuation and phase shifts attributed to the microwave circuit. Specifically, the attenuation is given by the constant $a$, the phase shift by $\beta$, and the electrical delay due to the finite length of the wiring by $\tau$. The expression within the square brackets describes the response of the resonator itself, including an additional correction $e^{i \phi}$ to describe potential asymmetries induced by spurious input- and output impedance or standing waves. We note that expanding $Q_{\mathrm{ext}}$ to the complex plane allows one to account for differences in impedance between the resonator and the feedline. In addition, $f$ and $f_{\mathrm{r}}$ denote the VNA driving frequency and resonator resonance frequency, respectively.
\section{Results and discussion}
\subsection{Comparison of NbTiN resonators grown on different substrates}
Representative raw transmission data for five resonators patterned on a NbTiN chip grown on highly resistive Si is presented as $|S_{21}(f)|^2$ in Fig.\,\ref{Fig: series2}(c) as open circles together with red lines representing fitting curves following Eq.\,(\ref{Eq: circle-fit}). The resonance frequencies $f_{\mathrm{r}}$ and quality factors $Q$ of the five resonators, as extracted from the fits, are listed in Tab.\,\ref{Tab: 1}.
Being interested mainly in the peak performance of our resonators, we choose to omit resonators with uncharacteristically low $Q_{\mathrm{int}}$ such as resonator R5 in Tab.\,\ref{Tab: 1} from our analysis. However, this only affects a single one of the 20 studied resonators. To provide a concise comparison of average resonator performance on different substrates, we fabricated and measured four sets of resonators and calculate the mean $Q$-factors. We account for the deviations between the on-chip resonators by setting the value for the net error of the $Q$-values as the sum of the fitting uncertainty and a statistical error $\sigma/\sqrt{N}$, where $\sigma$ is the standard deviation of the extracted $Q$-values and $N$ is the number of resonators. The extracted quality factors for all investigated NbTiN films grown on different substrates as well as the sample from the PTB are listed in Tab.\,\ref{Tab-summary}. The individual $Q$-factors of all resonators that comprise these mean values are listed in the SM section II \cite{Supplements}.
\begin{table}[htpb]
	\centering
	\begin{tabular}{|c|c|c|c|}
		\hline
		Substrate & $Q$ ($\times 10^3$) & $Q_{\mathrm{int}}$ ($\times 10^3$) & $Q_{\mathrm{ext}}$($\times 10^3$) \\ \hline
		$\mathrm{SiO_2}$ 	 &9.3$\pm$1.8 & 13.2$\pm$3.6 &38.7$\pm$10.7 \\ \hline
		$\mathrm{SiO_2}$ (PTB)	 &11.7$\pm$1.6 & 21.3$\pm$4.5 & 45.2$\pm$15.0 \\ \hline
		Si 	 &28.5$\pm$7.9 & 208.3$\pm$70.4 &37.8$\pm$11.6 \\ \hline
		Si + HF dip	 &74.6$\pm$7.0 & 182.5$\pm$22.7 & 126.3$\pm$9.8 \\ \hline
	\end{tabular}
	\caption{Mean fitted quality factors of the NbTiN resonators, deposited on different substrates. Measurements are performed at $T=2.2$ K and at a microwave power corresponding to $\simeq10^7$ photons on average in the resonator. }
	\label{Tab-summary}
\end{table}
For the WMI-NbTiN films grown on $\mathrm{SiO_2}$, we obtain an average internal quality factor of $Q_{\mathrm{int}}=(13.2\pm3.6)\cdot10^3$, which is about 1.5 times lower than for the PTB films [$Q_{\mathrm{int}}=(21.3\pm4.5)\cdot10^3$], despite the fact that these films have a larger surface roughness \cite{Supplements} and a lower superconducting transition temperature $T_{\mathrm{c}}=14$\,K. This result suggests that besides these parameters, the precise deposition conditions play an important role on the obtained quality factors. In particular, the selected deposition power $P$ controls the kinetic energy of the deposited target material clusters and hence the quality of the metal-substrate interface. For NbTiN grown on non-oxidized Si substrates, we achieve one order of magnitude higher average internal quality factors $Q_{\mathrm{int}}=(208.3\pm70.4)\cdot10^3$. We attribute this to increased losses from the metal-substrate interface induced by the diffusion of oxygen from the $\mathrm{SiO_2}$-layer into the NbTiN. In order to further reduce losses associated with defects or contaminations at the metal-substrate (MS) interface, we performed a buffered HF treatment to completely strip off the native oxide from one of the highly resistive Si substrates prior to deposition. However, for this sample we only achieved internal quality factors $Q_{\mathrm{int}}=(182.5\pm22.7)\cdot10^3$, comparable to those of the untreated Si-substrate. It seems that for the NbTiN resonators deposited on HF dipped Si, the reduced losses from the metal/substrate interface are compensated by increased losses related to the metal/air interface. This is corroborated by the larger RMS surface roughness for these films (see Fig.\,\ref{Fig: series1}). \\ Fitting the data to equation\,(\ref{Eq: circle-fit}) also allows to extract the external quality factor. Here, we find that all samples share an external $Q$-factor of $Q_{\mathrm{ext}}\approx4\cdot 10^4$ within the margin of error. This is expected as we used the identical layout for all chips with an equal spacing of $w_{\mathrm{gs}} = 40$ $\mathrm{\mu}$m between feed line and resonators, except for the HF dipped resonator chip. For the latter we used $w_{\mathrm{gs}} = 70$ $\mathrm{\mu}$m to avoid large uncertainties in $Q_{\mathrm{int}}$ for overcoupled resonators. We found $Q_{\mathrm{ext}}\approx1.3\cdot 10^5$. 
Finally, we note that the $Q_{\mathrm{int}}$ of our NbTiN resonators grown on Si are one order of magnitude higher than those of comparable resonators based on superconducting Nb described in previous publications \cite{Weichselbaumer2019, Zollitsch2015}. Therefore, our results represent a significant improvement in resonator quality.
\subsection{Performance of NbTiN resonators at elevated temperatures for ESR applications }
\begin{figure}[htpb]
	\centering
	\includegraphics[scale=1.0]{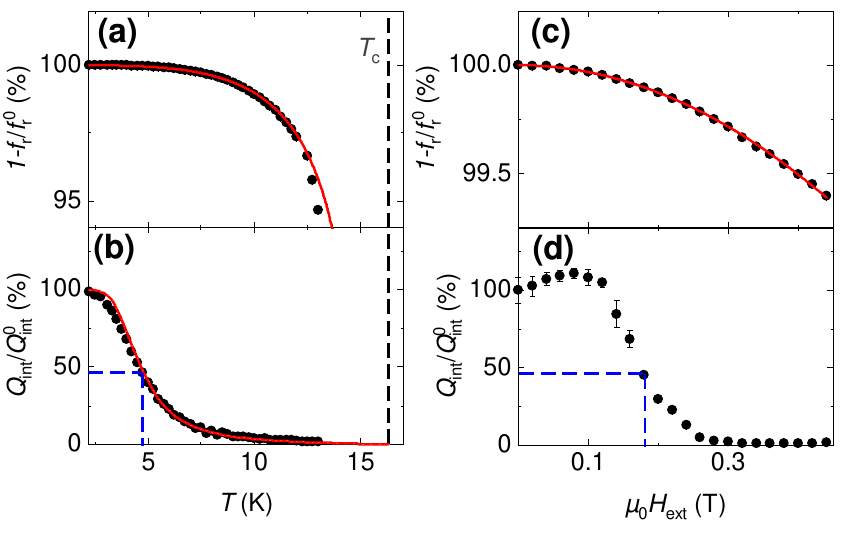}
	\caption{ Characteristic performance data of a NbTiN resonator (R1 from Tab.\,\ref{Tab: 1}) grown on a HF dipped, highly resistive Si substrate, at elevated temperatures and in an external magnetic field. (a) Normalized change in resonance frequency $1-\Delta f_{\mathrm{r}}(T)/ f_{\mathrm{r}}^0$ as a function of temperature $T$. The black dashed vertical line indicates $T_{\mathrm{c}}$ and the red curve represents a fit following the imaginary part of Eq.\,(\ref{Gl-Delta-fr-Modell}). (b) Internal quality factor $Q_{\mathrm{int}}$ normalized by its value at $T=2.2$ K. The red curve represents a fit to the real part of Eq.\,(\ref{Gl-Delta-fr-Modell}) to the data with a finite residual quality factor. (c) Normalized change in resonance frequency $1-\Delta f_{\mathrm{r}}(\mu_0H_{\mathrm{ext}})/ f_{\mathrm{r}}^0$ as a function of an external in-plane magnetic field $\mu_0H_{\mathrm{ext}}$. The red curve represents a quadratic fit. (d) Internal quality factor $Q_{\mathrm{int}}$ normalized by its value at $\mu_0H_{\mathrm{ext}}=0$ mT. The blue dashed lines in (b), (d) indicate where $Q_{\mathrm{int}}$ drops below $10^5$. }
	\label{Fig: series3}
\end{figure}

We next discuss the temperature dependence of the performance of the NbTiN-resonators. Doing so, we specifically focus on R1 from Tab.\,\ref{Tab: 1} grown on a HF dipped pristine silicon substrate. Figure\,\ref{Fig: series3} shows the relative change in the resonance frequency and the internal $Q$-factor in the temperature range $(2.2\leq T\leq13)$ K. 
As shown in Fig.\,\ref{Fig: series3}(a), the resonance frequency remains roughly constant in the range $T\leq7$ K, and then begins to visibly decrease with increasing $T$ approaching a relative change of 5.5\% at $T=13$ K. The internal quality factor plotted in Fig.\,\ref{Fig: series3}(b) shows a much stronger reduction with increasing $T$, approaching only about 1.5\% of its low temperature value at $T=13$ K. In general, the temperature dependence of the resonance frequency $f_\mathrm{r} (T)$ and the internal quality factor $Q_{\mathrm{int}} (T)$ of the resonators are determined by the temperature dependence of the surface reactance $X_\mathrm{s} (T)$ and surface resistance $R_\mathrm{s} (T)$ of the superconducting material. Following Refs. \cite{Minev2013, Reagor2013}, the relation between these quantities can be expressed as 
\begin{linenomath*}
	\begin{equation}
		\frac{1}{Q_{\mathrm{int}}(T)}+2i \frac{\Delta f_{\mathrm{r}}(T)}{f_{\mathrm{r}}^0}=\frac{\alpha}{\mu_0 \omega_{\mathrm{r}}^0\lambda_{\mathrm{L}}(0)}(R_{\mathrm{S}}(T)+i\delta X_{\mathrm{S}}(T)).
		\label{Gl-Delta-fr-Modell}
	\end{equation}
\end{linenomath*}
Here $\alpha$ denotes the contribution of the kinetic inductance to the total inductance, $\mu_0$ is the vacuum permeability and $\delta X_{\mathrm{S}}(T)=X_{\mathrm{S}}(T)-X_{\mathrm{S}}(0)$. For fitting the data, we use the expressions \cite{ tinkham2004introduction,Duzer1998}
\begin{linenomath*}
	\begin{align}
		\begin{aligned}
			R_{\mathrm{S}}(T)&=\frac{1}{2}\mu_0^2\lambda_{\mathrm{L}}^3(T)\omega^2\sigma_0\frac{n_{\mathrm{n}}(T)}{n}\\
			\delta X_{\mathrm{S}}(T)&=\mu_0\omega[\lambda_{\mathrm{L}}(T)-\lambda_{\mathrm{L}}(0)],
		\end{aligned}
	\end{align}
\end{linenomath*}
where $\sigma_0$ is the normal state conductivity, $n_\mathrm{n}(T)$ is the density of unpaired charge carriers and $n$ the total charge carrier density. For fitting, we determine $n_\mathrm{n}(T/T_{\mathrm{c}})/n=1-n_\mathrm{BCS}(T/T_{\mathrm{c}})/n$, where $n_\mathrm{BCS}$ is the superconducting particle density, according to BCS theory \cite{Einzel2003}. Equation\,(\ref{Gl-Delta-fr-Modell}) can be separated into an imaginary and a real part corresponding to the temperature evolution of the resonance frequency $f_{\mathrm{r}}$ and the quality factor $Q$. We make use of this and fit Eq.\,(\ref{Gl-Delta-fr-Modell}) to the data presented in Fig \ref{Fig: series3}(a) to extract $\alpha=0.124\pm0.001$. Here, we assume that $\lambda_\mathrm{L}(T)=\lambda_\mathrm{L}(0)/ \sqrt{1-(T/T_\mathrm{c})^4}$ \cite{Gorter1934}, with $\lambda_\mathrm{L}(0)=200$ nm \cite{Yu2005} and $T_{\mathrm{c}}= 16.3\pm0.1$ K, which was determined by recording $S_{21}(T)$ while raising the sample temperature (see section III of the SM \cite{Supplements}). This value of $\alpha$ falls into the expected range for planar 2D-resonators ($10^{-2}\leq\alpha\leq1$) \cite{Leduc2010a, Minev2013, Vayonakis2003}. We next fit the internal quality factor $Q_{\mathrm{int}}(T)$ in Fig.\,\ref{Fig: series3}(b) using the real part of Eq.\,(\ref{Gl-Delta-fr-Modell}) and extract a normal state conductivity of $\sigma_0=(11\pm4)\cdot 10^6\;\Omega^{-1}\mathrm{m}^{-1}$, considering the absolute $Q_{\mathrm{int}}$. This result is in decent agreement with the results of Van-der-Pauw transport experiments \cite{VanderPAUW1991a} on our blanketed films, where we detected varying $\sigma_0$ in the low $10^6\;\Omega^{-1}\mathrm{m}^{-1}$-range. We also experimentally measured the microwave transmission for $T>13$ K. However, here the increasingly steep frequency dependence with temperature limits the reliable determination of the internal $Q$-factors, as frequency fluctuations start to dominate the linewidth.\\ We further investigate the performance of our resonators in an external in-plane magnetic field for ESR- and FMR-applications [see Fig.\,\ref{Fig: series1}(b)]. In Fig.\,\ref{Fig: series3}(c) and (d), we plot the normalized shift in resonance frequency $1-\Delta f_{\mathrm{r}}(T)/ f_{\mathrm{r}}^0$ and internal quality factor $Q_{\mathrm{int}}$ as function of applied external in-plane magnetic field $\mu_0H_{\mathrm{ext}}$ in the range from (0-440) mT. For higher $\mu_0H_{\mathrm{ext}}$, the fitting of the $S_{21}(f)$-curves becomes unreliable. The observed reduction in $f_{\mathrm{r}}$ is fitted with the common quadratic relation [$\Delta f_{\mathrm{r}}(\mu_0H_{\mathrm{ext}})\propto (\mu_0H_{\mathrm{ext}})^2$] \cite{Zollitsch2019, Xu2019}. For the internal quality factor $Q_{\mathrm{int}}$ in Fig.\,\ref{Fig: series3}(d), we observe a nearly constant $Q_{\mathrm{int}}$ until approximately $\mu_0H_{\mathrm{ext}}=130$ mT and a strong decrease for the internal quality factor for higher $\mu_0H_{\mathrm{ext}}$. 
Regarding the performance of our resonators at elevated temperatures and in an external field, we achieve a $Q_{\mathrm{int}}>10^5$ for temperatures $T\leq4.8$ K and external in-plane fields $\mu_0H_{\mathrm{ext}}\leq180$ mT as indicated by the blue dashed lines in Fig.\,\ref{Fig: series3}(b) and (d). Given the high field- and temperature-stability demonstrated in these measurements, our NbTiN resonators appear to be well suited for ESR and FMR experiments for $T\leq13$ K and $\mu_0H_{\mathrm{ext}} \leq440$ mT.
\subsection{Performance of NbTiN resonators at mK temperatures for quantum science applications}
\begin{figure}[htpb]
	\centering
	\includegraphics[scale=1.0]{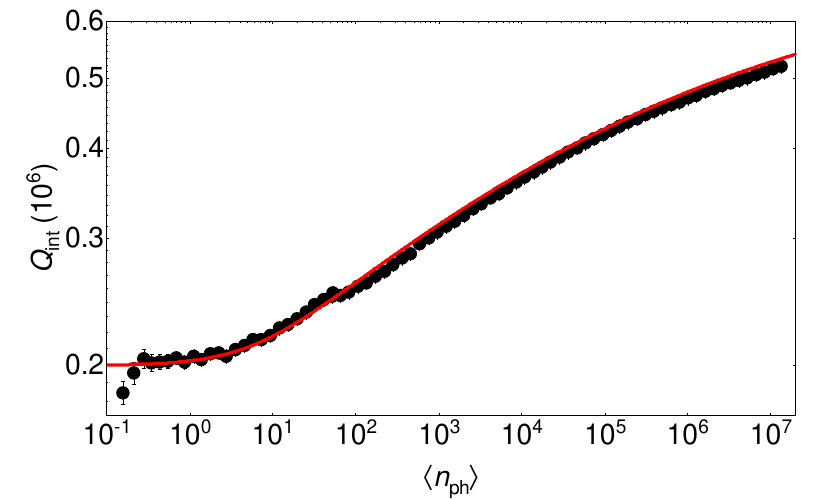}
	\caption{Intrinsic quality factor of a NbTiN resonator grown on HF treated Si (R1 from Tab.\,\ref{Tab: 1}) and measured in a dry dilution refrigerator at $T<7$ mK as function of $\langle n_{\mathrm{ph}}\rangle$. The $Q$-factors and $f_{\mathrm{r}}$ are extracted by fitting the $S_{21}(f)$-spectra to Eq.\,(\ref{Eq: circle-fit}) and are then used to convert the calibrated input power $P$ to $\langle n_{\mathrm{ph}}\rangle$ using Eq.\,(\ref{Eq: nph}). The red line represents a fit to Eq.\,(\ref{Eq: Qint-nph}). }
	\label{Fig: series4}
\end{figure}

To investigate the performance of our resonators for quantum applications, we investigate the complex microwave transmission in a dry dilution refrigerator with a base temperature of $<7$ mK (see Fig.\,S3 in section IV of the SM for the microwave setup \cite{Supplements}). Using the same sample package as above, we determine $Q_{\mathrm{int}}$ as function of the microwave spectroscopy power $P$, which is translated into an average photon occupancy $\langle n_{\mathrm{ph}}\rangle$ employing additional calibration experiments. Figure\,\ref{Fig: series4} displays the internal quality factor of resonator R1 from Tab.\,\ref{Tab: 1} grown on HF dipped, high-resistivity Si as function of the average intra-resonator photon number $\langle n_{\mathrm{ph}}\rangle$, which is related to $P$ via 
\begin{equation}
	\langle n_{\mathrm{ph}}\rangle=P\left(\frac{1}{h\cdot f_{\mathrm{r}}}\cdot\frac{2\kappa_{\mathrm{ext}}/2}{\kappa/2}\right).
	\label{Eq: nph}
\end{equation}
In Fig.\,\ref{Fig: series4}, we observe a continuous increase in $Q_{\mathrm{int}}$ with increasing resonator photon population from $\approx$ $2\cdot10^5$ at $\langle n_{\mathrm{ph}}\rangle\simeq1$ to $\approx$ $5.2\cdot10^5$ at $\langle n_{\mathrm{ph}}\rangle\simeq10^7$. Additionally, we observe a plateau at $Q_{\mathrm{int}}\simeq2\cdot 10^5$ at low $\langle n_{\mathrm{ph}}\rangle=10^{-1}$-$10^0$ representing the single photon regime. However, a saturation in $Q_{\mathrm{int}}$ at high average photon numbers $\langle n_{\mathrm{ph}}\rangle$ is not observed, suggesting that TLS induced losses still constitute a substantial loss channel even at these high $\langle n_{\mathrm{ph}}\rangle$. This behavior has already been reported for NbTiN in Refs. \cite{Bruno2015, Barends2010a}. The net $Q_{\mathrm{int}}$ at millikelvin temperatures is about a factor of two higher compared to the cryogenic setup at $T=2.2$ K, as magnetic field and infrared shielding reduce losses due to trapped vortices \cite{Song2009} and infrared radiation \cite{Serniak2018}, while QP losses are frozen out \cite{Zmuidzinas2012}.
We fit the measured power dependence of $Q_{\mathrm{int}}(\langle n_{\mathrm{ph}}\rangle)$ with \cite{Earnest2018, Burnett2018,Richardson2016, Goetz2016}
\begin{equation}
	\frac{1}{Q_{\mathrm{int}}(\langle n_{\mathrm{ph}}\rangle)}=F\,\mathrm{tan}\,\delta_{\mathrm{TLS}}^0\left(1+\frac{\langle n_{\mathrm{ph}}\rangle}{ \langle n_{\mathrm{ph}}^c\rangle}\right)^{-\alpha}+\frac{1}{Q^*},
	\label{Eq: Qint-nph}
\end{equation}
where $F\,\mathrm{tan}\,\delta_{\mathrm{TLS}}^0=1/Q_{\mathrm{TLS}}$ represents the total TLS loss at $\langle n_{\mathrm{ph}}\rangle=0$ and $T=0$ K, $\langle n_{\mathrm{ph}}^c\rangle$ is a critical mean photon number above which the TLS defects are saturated and $\alpha$ is an exponent indicating the deviation from the standard TLS model (in the standard model $\alpha=0.5$). Finally, the offset $1/Q^*$ accounts for all other (non-TLS associated) loss channels.
From the fit of Eq.\,(\ref{Eq: Qint-nph}) to $Q_{\mathrm{int}}$ in Fig.\,\ref{Fig: series4}, we extract the total TLS loss $F\,\mathrm{tan}\,\delta_{\mathrm{TLS}}^0=(3.6\pm0.1)\cdot10^{-6}$, $\langle n_{\mathrm{ph}}^c\rangle=(6.1\pm0.7)\cdot10^3$, $\alpha=0.14\pm0.03$ and $Q^*=(7.2\pm0.2)\cdot 10^5$. The value of $\alpha$ deviates from $\alpha=0.5$ expected within the conventional TLS model, however, it is in agreement with the values found for Al \cite{Earnest2018}. The value for the total TLS loss consists of contributions originating from the dielectric loss at the metal/air, metal/substrate and substrate/air interface as well as the impact of the loss tangent of the substrate (S). Assuming that these are known factors and accounting for the geometry of the resonator, one can compute $F\,\mathrm{tan}\,\delta_{\mathrm{TLS}}^0$ using \textit{Ansys HFSS (Ansys\SymbReg Electronics Desktop 2020 R2)} finite element simulation software (for details see SM Sec. V \cite{Supplements}). Using the loss tangents $\mathrm{tan}(\delta_{\mathrm{i}})$, layer thicknesses $d_{\mathrm{i}}$ and dielectric constants $\epsilon_{\mathrm{i}}/\epsilon_{0}$ of the four dielectric regions from Ref. \cite{Melville2020}, we expect $F\,\mathrm{tan}\delta_{\mathrm{TLS}}^0\approx0.5\cdot10^{-6}$. This value is approximately one order of magnitude smaller than the experimental value. This suggests that either the NbTiN/Si interface is not well described within our model or that other contributions e.g. originating from the metal/air and metal/substrate interface play a more dominant role. A detailed study disentangling and identifying the source of this discrepancy is beyond the scope of this work. The large value of $\langle n_{\mathrm{ph}}^c\rangle$ extracted from this fit is a consequence of the continuous rise in $Q_{\mathrm{int}}$ with $\langle n_{\mathrm{ph}}\rangle$ as also previously observed for NbTiN in Refs. \cite{Bruno2015, Barends2010a}, but also for other materials i.e. elementary Nb\cite{Verjauw2021, Altoe2020}, NbN \cite{Zollitsch2019, Yu2021} and TiN \cite{Melville2020}. One potential explanation is that NbTiN is considered an extreme 'dirty superconductor' with a large defect density. Interestingly, the $Q_{\mathrm{int}}$ observed in our resonators is comparable to the resonators presented in Ref. \cite{Barends2010a} and to NbTiN resonators, which have not been treated with hexamethyldisilazane (HMDS) prior deposition and patterned with deep reactive ion etching (DRIE) in Ref. \cite{Bruno2015}. While this demonstrates the state-of-the-art performance of our NbTiN films used in high-$Q$ microwave resonators, it also suggests that defects within the material might still be a limiting factor. In future studies, substrate surface passivation treatments such as HMDS-based recipied to reduce oxygen-related TLS at the metal/substrate-interface need to be pursued \cite{Bruno2015, Earnest2018}. To address and improve on these loss channels, future investigations of the impact on stoichiometry of NbTiN films as well as crystalline growth modes may be beneficial. 
\section{Summary}
In summary, we fabricated planar microwave resonators out of NbTiN thin films, which were grown on thermally oxidized Si and highly resistive Si (001) substrates and we measured and compared their internal quality factors $Q_{\mathrm{int}}$ at $T=2.2$ K. We find $Q_{\mathrm{int}}\simeq2\cdot 10^5$ for resonators grown on Si-substrates, regardless of a HF treatment prior to deposition and $Q_{\mathrm{int}}\simeq 10^4$ for resonators grown on $\mathrm{SiO_2}$. A comparison of our results to a resonator fabricated from a NbTiN thin film grown on $\mathrm{SiO_2}$ at the \textit{Physikalisch-Technischen Bundesanstalt} (PTB) revealed comparable $Q_{\mathrm{int}}$ deviating only by a factor of about 1.5. The $Q_{\mathrm{int}}$ for NbTiN grown on Si are one order of magnitude higher than those reported in our previous publications for comparable resonator layouts based on superconducting Nb \cite{Weichselbaumer2019, Zollitsch2015}. With the focus on ESR and FMR applications, we demonstrated the usability of our resonators at elevated temperatures of up to $T=13$ K and magnetic in-plane fields up to $\mu_0H_{\mathrm{ext}}=440$ mT with a $Q_{\mathrm{int}}>10^5$ for $T\leq4.8$ K and $\mu_0H_{\mathrm{ext}}\leq180$ mT. Regarding their integration in Qubit devices i.e. at millikelvin temperatures, our NbTiN resonators exhibit $Q_{\mathrm{int}}\approx5\cdot10^5$ in the high power limit and $Q_{\mathrm{int}}\approx2\cdot10^5$ in the single photon limit. 
\section{Acknowledgments}
We acknowledge financial support by the German
Research Foundation (DFG, Deutsche Forschungsgemeinschaft) via Germany’s Excellence Strategy EXC-2111-390814868 and the Munich Quantum Valley, which is supported by the Bavarian state government with funds from the Hightech Agenda Bayern Plus.\\
The work at PTB work was co-funded by the Deutsche Forschungsgemeinschaft (DFG, German Research Foundation) under Germany’s Excellence Strategy – EXC-2123 QuantumFrontiers – 390837967.\\
 Furthermore, we want to thank Sebastian Kammerer and Niklas Brukmoser for performing the HF surface treatment as well as Yuki Nojiri and Kedar Honasoge for their support with the mK-experiments.



\section*{References}
\bibliography{library}

\end{document}